\documentclass
[twocolumn,showpacs,preprintnumbers,amsmath,amssymb,prc]{revtex4}
\usepackage{graphicx}
\usepackage{dcolumn}
\usepackage{bm}

\begin{document}
\title{Liquid gas phase transition in hypernuclei}

\author{S. Mallik and G. Chaudhuri}

\affiliation{Theoretical Physics Division, Variable Energy Cyclotron Centre, 1/AF Bidhan Nagar, Kolkata 700064, India}

\begin{abstract}
The fragmentation of excited hypernuclear system formed in heavy ion collisions has been described by the canonical thermodynamical model extended
to three component systems. The multiplicity distribution of the fragments has been analyzed in detail and it has been observed that the hyperons have the tendency to get attached to the heavier fragments. Another important observation is the phase coexistence of the hyperons, a phenomenon which is linked to
liquid gas phase transition in strange matter.
\end{abstract}

\pacs{25.70.Mn, 25.70.Pq}

\maketitle
{\bf {\it Introduction:-}} The physics of hypernuclei is an important area of study in the regime of high energy heavy ion collisions. It has been observed that baryons and mesons(strange hadrons) are produced abundantly in high energy heavy ion reactions. Hypernuclei are formed when the strange hyperons or baryons are captured by the nuclei. The $\Lambda$-nucleon interactions are well studied and the potential depth of $\Lambda$ hyperons is such that bound $\Lambda$ hypernuclear states exist. The only bound $\Sigma$ hypernucleus known so far is which is bound by isospin forces \cite{Millener}.  Several $\Xi$ hypernuclear states are reported in the literature and hence its interaction with nucleon seems to be attractive.  On the other hand the hyperon-hyperon interaction is not really well known; a few double $\Lambda$ hypernuclear states have been reported. The interaction between other pairs of hyperons as $\Lambda$ $\Xi$ or $\Xi$ $\Xi$ is not known experimentally \cite{Bielich1}. The formation of multi-strange nuclei is especially important in order to study the properties of strange matter. Deep understanding of strange matter is extremely important for the formulation of  models of strong interaction \cite{Papaz}. Another important application  of this study is the core of neutron stars \cite{Bielich} where the hyperons are expected to be produced in abundance  at high density nuclear matter. The stability of hypernuclei beyond the neutron and proton driplines(normal nuclear chart) is also a fascinating subject which is important in recent day activities \cite{Botvina1,Botvina2,Botvina3}. The knowledge of structure of normal nuclei \cite{Hashimoto} as well as the extension of the nuclear chart into the strangeness sector \cite{Schaffner,Greiner,Samanta} gets valuable input from the results of hypernuclei study. Another important area in the study of intermediate energy heavy ion collisions is the phenomenon of phase coexistence or liquid gas phase transition \cite{Siemens,Pochodzalla,Dasgupta_Phase_transition}. The appearance of 'liquid-like' as well as 'gas-like' fragments simultaneously over a temperature interval is linked to first order phase transition. Whether this phase-coexistence will still persist in the presence of hyper-fragments(strange fragments) is the object of investigation in this work.\\
The canonical thermodynamical model has already been extended to three component systems \cite{Dasgupta_hyperon1} i.e, inclusion of hyperons(usually $\Lambda$ in addition to the neutrons and the protons). Due to fragmentation of the PLF, normal (non-strange) components as well as hypernuclei will be formed. In previous works \cite{Dasgupta_hyperon1, Dasgupta_hyperon2}, the total number of strange particles was confined to 2. In this work we include the possibility of existence of multiple (more than 2) strange particles.\\
In a recent paper, \cite{Dasgupta_hyperon2} a hybrid model based on participant-spectator picture combined with the Canonical
Thermodynamical Model(CTM) model has been used to determine the production cross section
of  a hypernucleus in high energy heavy-ion collisions,. For heavy ion collisions in 3-10 GeV range, the following scenario
 (backed by experiments) is applicable.  For a general impact parameter,
 there is a region of violent collision called the participating region. In addition there is a mildly excited projectile like fragment(PLF)
 and also a mildly excited target like fragment(TLF). Physics of both PLF and TLF are similar for symmetric collisions; here we concentrate
our analysis on PLF.
Because of excitation energy (usually characterized by a temperature, T) PLF will break up into many fragments \cite{Mallik2,Mallik3,Mallik9}
 and the velocities of the fragments are centered around the velocity of the projectile. In fixed target experiments they are emitted in a forward cone and are more easily recognizable. In the participating region, apart from original neutrons and protons, particles (pions, $\Lambda$'s, etc.) are produced. The produced $\Lambda$'s have an extended rapidity range. Those produced in the rapidity range close to that of the projectile and having total momenta in the PLF frame up to the Fermi momenta can be trapped in the PLF and form hypernuclei \cite{Wakai,Saito,Gaitanos}. At higher energies multiple hyperons can get attached to the PLF. In this work we consider a maximum number of eight hyperons being attached to the PLF. The fragmentation of the PLF into different composites(strange and non-strange) is calculated using the three component CTM \cite{Dasgupta_hyperon1,Dasgupta_hyperon2}.\\
The main motivation of this work is to analyze the composition of the fragments produced from fragmentation of PLF
 which initially has multiple hyperons attached to it. The important feature which emerges from the results is that hyperons have
 greater affinity of getting attached to the higher mass fragments. The most striking feature of the distribution of the hyperons is the
  phase coexistence, a feature which has already been  observed in the case of normal(non-strange) fragments \cite{Siemens,Pochodzalla,Dasgupta_Phase_transition}. The typical 'U' shaped distribution  observed in the fragmentation of non-strange(normal) nuclei is also exhibited by the fragmentation of the strange nuclei irrespective of the amount of strangeness content. One can infer that phase transition which is a characteristic feature of fragmentation of  normal nuclei also persists  in the case of hyperfragments.\\

{\bf {\it Theoretical formalism:-}} The Canonical Thermodynamical Model (CTM) for two kinds of particles (neutron and proton) is well-known and has had long usage \cite{Das1}. This has been extended to three kinds of particles (neutron, proton and $\Lambda$) few years back
 \cite{Dasgupta_hyperon1,Dasgupta_hyperon2}. In this section, the 3-component Canonical Thermodynamical Model is discussed briefly. Assuming
that a system with $A_0$ baryons, $Z_0$ protons and $H_0$ hyperons at temperature $T$, has expanded to a higher than normal volume, the partitioning into different composites can be calculated according to the rules of equilibrium statistical mechanics. The canonical partition function is given by
\begin{equation}
Q_{A_0,Z_0,H_0}=\sum\prod\frac{(\omega_{a,z,h})^{n_{a,z,h}}}{n_{a,z,h}!}
\end{equation}
Here the product is over all fragments of one break up channel and sum is over all possible channels of break-up (the number of such channel
is enormous); $\omega_{a,z,h}$ is the partition function of one composite with $a$ baryons $z$ protons and $h$ hyperons whereas
 $n_{a,z,h}$ is the number of this composite in the given channel. The one-body partition function $\omega_{a,z,h}$ is a product of two parts: one arising from the translational motion and another is the intrinsic partition function of the composite:
\begin{equation}
\omega_{a,z,h}=\frac{V}{h^3}(2\pi T)^{3/2}\{(a-h)m_n+hm_h\}^{3/2}\times z_{a,z,h}(int)
\end{equation}
Here $m_n$ and $m_h$ are masses of nucleon (we use 938 MeV) and hyperon (we use 1116 MeV for $\Lambda$ hyperon) respectively.
$V$ is the volume available for translational motion; $V$ will be less than $V_f$, the volume to which the system has expanded at break up.
 We use $V = V_f - V_0$ , where $V_0$ is the normal nuclear volume. Since hyperfragments are generally studied from PLF, hence we have
considered $V_f = 3V_0$.
The average number of composites with $a$ baryons, $z$ protons and $h$ hyperons can be written as
\begin{equation}
<n_{a,z,h}>=\frac{\omega_{a,z,h}Q_{A_0-a,Z_0-z,H_0-h}}{Q_{A_0,Z_0,H_0}}
\end{equation}
Each allowed break up channel in Eq. 1 must satisfy, total baryon, proton and hyperon conservation i.e.
\begin{eqnarray}
\sum an_{a,z,h} &=& A_0  \nonumber \\
\sum zn_{a,z,h} &=& Z_0  \nonumber  \\
\sum hn_{a,z,h} &=& H_0
\end{eqnarray}
Substituting Eq.(3) in these three constraint conditions, three different recursion relations \cite{Chase} can be obtained.
 Any one recursion relation can be used for calculating $Q_{A_0,Z_0,H_0}$. For example
\begin{equation}
Q_{A_0,Z_0,H_0}=\frac{1}{A_0}\sum_{a,z,h}a\omega_{a,z,h}Q_{A_0-a,Z_0-z,H_0-h}
\end{equation}
Therefore calculation of  any partition function  using this recursion relation will require very short computational time and
then substituting  those in Eq. 3 one can calculate the average multiplicity $\langle n_{a,z,h}\rangle$ easily.\\
To construct $z_{int}(a,z,h)$, experimental binding energies are used for low mass nuclei and hypernuclei, and for higher
 masses a liquid drop formula is used. The neutron, proton and $\Lambda$ particles are taken as the fundamental blocks
 therefore $z_{int}(1,0,0)$=$z_{int}(1,1,0)$=$z_{int}(1,0,1)$=1. For deuteron, triton, $^3$He and $^4$He
we use $z_{a,z,0}(int)=(2s_{a,z,0}+1)\exp(-\beta e_{a,z,0}(gr))$ where $\beta=1/T, E_{I,J}(gr)$ is the ground state energy
(taken from experimental data) and $(2s_{I,J}+1)$ is the experimental spin degeneracy of the ground state. For $1<a\le8$, the ground state binding energies and excited state energies are taken from experimental data \cite{Dasgupta_hyperon2}. For heavier nuclei and hypernuclei, liquid-drop formula is used for calculating ground state energy \cite{Botvina3}. This is given by
\begin{eqnarray}
e_{a,z,h}(gr)=-16a+\sigma(T)a^{2/3}+0.72kz^2/(a^{1/3})\nonumber\\
+25(a-h-2z)^2/(a-h)
-10.68h+21.27h/(a^{1/3})
\end{eqnarray}

where $\sigma(T)$ is the surface tension which is given by $\sigma(T)=\sigma_{0}\{(T_{c}^2-T^2)/(T_{c}^2+T^2)\}^{5/4}$ with $\sigma_{0}=18.0$
MeV and $T_{c}=18.0$ MeV and k is the correction factor in Coulomb energy which incorporates the effect of its long-range behavior
 by Wigner-Seitz approximation as in Ref. \cite{Bondorf1}. We include all nuclei within drip lines in constructing
 the partition function. Another useful parametrization in liquid drop formula for hypernuclei was proposed
by Samanta et. al. \cite{Samanta}. A comparative study of these two formula in the case of hyperfragmentation was described
in Ref. \cite{Botvina3} and finally the one used here was chosen because it produces results closer to the experimental data.\\
The study of the liquid-drop model formula (which has been used in our model), reveals that by adding hyperons the stability of the fragments increase for mass numbers $a>8$. Hence this implies that the hyperon-nucleon interaction is attractive for this mass range. For $a\le8$, this liquid-drop formula is not suitable and so we have used experimental binding energies for these lower mass nuclei or hypernuclei. It is known that, $^4$H or $^5$He are not stable, but when one $\Lambda$ is added the corresponding nuclei ($^4_{\Lambda}$H or $^5_{\Lambda}$He) become stable. Hence this establishes the attractive nature of the hyperon-nucleon interaction. However from this work, it is difficult to comment about hyperon-hyperon interaction as here it is not possible to isolate it from the other two interactions (nucleon-nucleon and nucleon-hyperon).\\
In addition to the liquid-drop formula we have also included the contribution to $z_{int}(a,z,h)$ due from the excited states. This gives a
multiplicative factor $exp (r(T)Ta/\epsilon_0)$ where we have introduced a correction term $r(T)=\frac{12}{12+T}$ to the expression used in Ref. \cite{Bondorf1}. This slows down the increase of $z_{int}(a,z,h)$ due to excited states as $T$ increases.\\

{\bf {\it Results and Discussions:-}}We  have computed the average number  of normal and hyperfragments of different mass, charge and strangeness by
 the canonical thermodynamical model. The fragmenting hypernuclei is assumed to have mass number $A_0 = 128$, charge $Z_0= 50$ and
total strangeness $H_0=8$. We have calculated the strangeness distribution of the hyperfragments
 in a very much similar way one calculates the charge or mass distribution of the fragments.
 Fig. 1 shows this distribution of hyperfragments ($\langle n_h\rangle=\sum_{a,z}\langle n_{a,z,h}\rangle$) at different
temperatures (excitation energies). At lowest temperature $3$ MeV, the distribution resembles 'U' shape.
  This nature is very much similar to what one obtains in the case of mass distribution
of normal fragments at low temperature. This 'U' shape of mass distribution for normal fragments and the lowering down of the
height of the maxima on the higher mass side as temperature is increased
 is usually linked to first-order phase transition or phase coexsistence\cite{Das1,Dasgupta_Phase_transition,Chaudhuri1,Chaudhuri2}. Similar feature
also emerges in the case of strange fragments or hyperfragments. With similar reasoning as in the case of normal (non-strange)
 fragments, we can associate this phenomenon in hyperfragments (see Fig. 1) with phase coexistence or liquid-gas phase transition.
 There is existence of hyperfragments with small strangeness content
\begin{figure}[t]
\includegraphics[width=5.2cm,keepaspectratio=true]{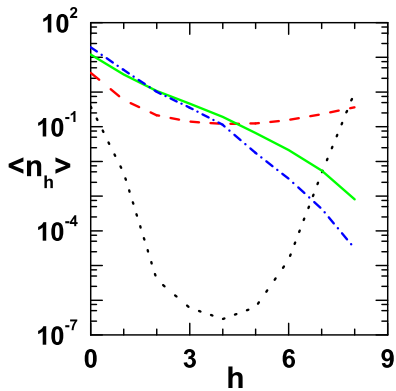}
\caption{(Color online) Distribution of hyperfragments produced from the fragmentation of $A_0=128$, $Z_0=50$, $H_0=8$ at $T=$ 3 MeV (black dotted line), 5 MeV (red dashed line), 7 MeV (green solid line) and 10 MeV (blue dash-dotted line).}
\end{figure}
\begin{figure}[h]
\includegraphics[width=\columnwidth,keepaspectratio=true]{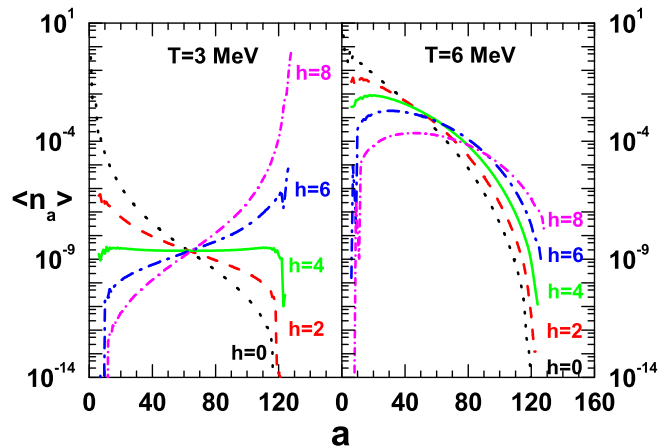}
\caption{(Color online) Mass distribution h=0, 2, 4, 6 and 8 hyperfragments produced from the fragmentation of $A_0=128$, $Z_0=50$, $H_0=8$ at $T=$ 3 MeV (left panel) and 6 MeV (right panel).}
\end{figure}
\begin{figure}[b]
\includegraphics[width=6.5cm,keepaspectratio=true]{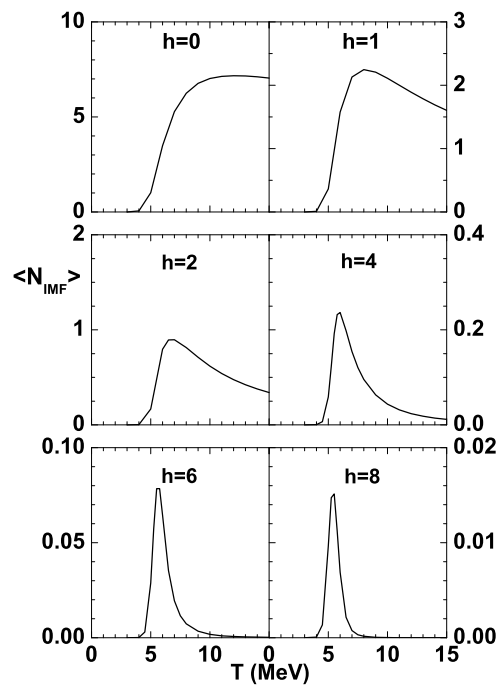}
\caption{Temperature dependence of intermediate mass fragments containing different strangeness produced from the fragmentation of $A_0=128$, $Z_0=50$, $H_0=8$.}
\end{figure}
as well as large strangeness content at the same time.
This $\langle n_h\rangle$ vs $h$ plot is similar to $\langle n_a\rangle$ vs $a$ plot \cite{Das1}
at different temperatures. As we increase the temperature,
the so called 'U' shape gradually flattens and finally at higher temperature, it changes to monotonically decreasing pattern as is seen from the figure. This can be inferred as disappearing of one phase as the temperature is increased. Though the difference in strangeness content between the two phases is not very much in the present case, but still we can refer to this pattern as phase coexistence in the
hyperfragments. This calculation is confined to a maximum number of $8$
 hyperons but we believe that if it is extended to larger number of hyperons,
the pattern will remain same and will confirm our inference from this figure.\\
In order to further analyze the distribution of strangeness content in different fragments of varying mass,  we have  calculated their mass
 distribution with different $h$ values separately. Fig. 2 displays this multiplicity distribution at two different temperatures $3$ and $6$ MeV. First let us concentrate on the lower temperature , that is $3$ MeV. For $h=0$ that is  for the normal fragments with no strangeness , the nature of the curve is monotonically decreasing which shows that cross-section of formation of heavier fragments with no strangeness content is extremely less. This can be interpreted by the fact that
the hyperons tend to get attached to the heavier fragments at lower temperature and hence most of the heavier fragments are strange.
This is confirmed by the other plots in the same figure which shows the mass distribution for fragments with different strangeness
 content, i.e, $h=2,4,6$ or $8$. More is the mass number of a fragment, greater is the probability of more hyperons getting attached to it.
 On the contrary, the strangeness content of lower mass fragments is comparatively less. The multiplicity of fragments with $h = 0$ or $h = 2$ is much more for lower values of $a$. For small strangeness content, the multiplicity decreases as one increases $a$. The right side of this figure shows the similar plot for a higher temperature $T=6$ MeV. As the temperature increases, fragments with higher mass decrease for obvious reasons. Lighter mass fragments are predominant at higher temperatures and they contain little or no strangeness.\\
\begin{figure}[h]
\includegraphics[width=6.5cm,keepaspectratio=true]{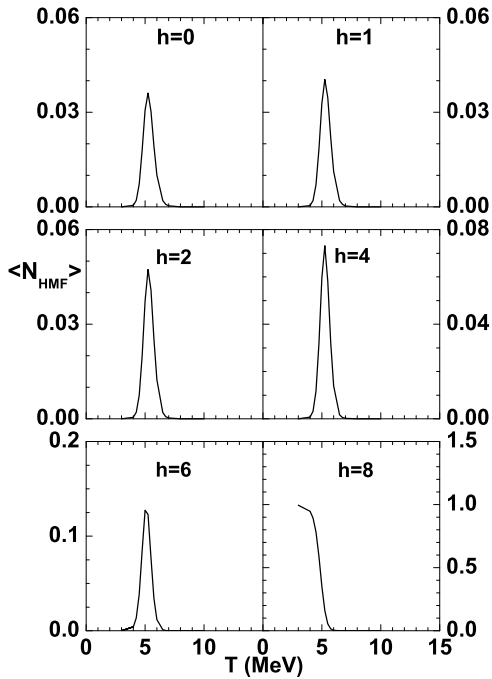}
\caption{Temperature dependence of higher mass fragments containing different strangeness produced from the fragmentation of $A_0=128$, $Z_0=50$, $H_0=8$.}
\end{figure}
The "rise and fall" nature of intermediate mass fragment (IMF) multiplicities is also an important signature of liquid gas phase
transition for normal nuclei \cite{Dasgupta_Phase_transition,Peaslee,Ogilvie,Tsang}. In this article, our aim is to investigate
how IMF and HMF (heavier mass fragment) multiplicities change with temperature for hypernuclei with different strangeness content.
Fig. 3 shows the variation of the  average number of intermediate mass
 fragments [$\langle n_{IMF}(h)\rangle=\sum_{z=3}^{20}\langle n_{a,z,h}\rangle$] with temperature for
 different $h$ content. For $h = 0$, $\langle N_{IMF}\rangle$ increases with $T$. This implies that the multiplicity of
 ordinary intermediate mass fragments increase monotonically with temperature. For $h=1$ or for higher values of $h$, the multiplicity
first increases, reaches a peak at a certain temperature and then decreases. Though the trend is similar for different $h$ values,
 the exact nature of the variation is different. The multiplicity at higher temperatures is more for fragments with lesser
 strangeness content, i.e, with lower values of $h$. Naturally, more strange is the fragment, less is the multiplicity of IMF which
once again establishes the tendency of hyperons to get  preferentially attached to the heavier mass fragments.\\
Fig. 4 shows the variation of multiplicity of heavier mass fragments ($\langle n_{HMF}(h)\rangle=\sum_{z\rangle {20}}\langle n_{a,z,h}\rangle$)with temperature for different $h$ values. Since the heavier mass fragments are predominantly strange, hence they have maximum multiplicity for $h$ =8. This can be easily understood refereing to fig. 2(left panel).\\
\begin{figure}[h]
\includegraphics[width=5.2cm,keepaspectratio=true]{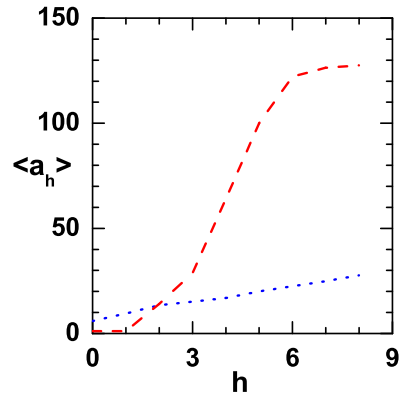}
\caption{(Color online) Average mass of the fragments ($\langle a_h\rangle$) with different strangeness ($h$) due to fragmentation of $A_0=128$, $Z_0=50$, $H_0=8$ at $T=3$ MeV (red dashed line) and $7$ MeV (blue dotted line).}
\end{figure}
Fig 5. shows the variation of $\langle a_h\rangle$($=\sum_{a,z}a\langle n_{a,z,h}\rangle/\sum_{a,z}\langle n_{a,z,h}\rangle$) with $h$ for two different temperatures. At the lower temperature $3$ MeV, the steep
increase of $\langle a_h\rangle$ with $h$ signifies once again the tendency attachment of more number of hyperons to the heavier fragments.
At lower excitation energy(temperature), formation of heavier fragments is dominant and that is  being reflected in the plot. Average mass of ordinary fragments(with no strangeness) is much less as compared to the strange ones. But this feature drastically changes at higher temperature where the variation of $\langle a_h\rangle$ vs h is much flatter. This is mainly due to the fact that heavier mass fragments are dominant at lower temperature and their formation is far less probable as one increases temperature. The average value of $a_h$ for $h$ =8 is about $5$ times more in case of $3$ MeV than $7$ MeV.\\
\begin{figure}[t]
\includegraphics[width=5.2cm,keepaspectratio=true]{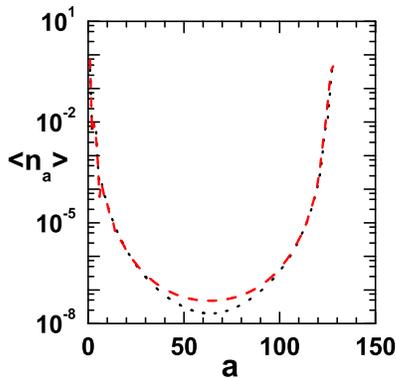}
\caption{(Color online) Mass distribution of hyperfragments (and/or fragments) produced at $T=$ 3 MeV from the fragmentation of two different sources having same $A_0=128$, $Z_0=50$ but different $H_0=8$ (black dotted line) and $H_0=0$ (red dashed line).}
\end{figure}
Fig 6. shows the variation of $\langle n_a\rangle$ with mass number $a$(mass distribution) for fragmentation of nuclei
 with $H$=0 and $H$ = 8 at $T$ = 3 MeV. The important feature is that both the curves are very similar in nature.
 If we concentrate on the fragmentation of the ordinary nuclei with no strangeness, the mass distribution
 displays an 'U' shaped variation which is expected at lower temperature (3 MeV). This shape gradually disappears as the
 temperature is increased. This feature indicates liquid gas phase transition or phase co-existence i.e. existence
 of 'liquid-like'(heavier) and gas-like' (lighter) fragments. This phenomenon has been well studied
for non strange fragments in both statistical \cite{Dasgupta_Phase_transition,Das1,Chaudhuri1,Chaudhuri2} and dynamical \cite{Mallik10} models as well as in experimental observations \cite{Pochodzalla,Borderie} and hence we will not elaborate here. Our main motivation is to
investigate the fragmentation of a nucleus with considerable amount of strangeness $H$ = 8.
It is quite amazing that the nature of mass distribution is similar and the two curves are pretty close to each other.
This establishes the fact that the first order phase transition (co-existence) still persists in the presence of hyperfragments.
This feature is independent of the strangeness content of the fragments.\\

{\bf {\it Summary and Conclusion:-}}
The fragmentation of a nucleus with multiple hyperons attached to it has been studied with the motivation to analyze the fragmentation pattern. The results clearly point to the affinity of the hyperons getting preferentially attached to the higher mass (heavier) fragments. Another important feature which emerges from the mass distribution is coexistence of liquid-like and gas-like hyper-fragments in a certain temperature interval. This phase coexistence is indicative of first order phase transition occurring in the fragmentation of nuclei with multiple hyperons. Above the transition temperature, the heavier fragments disappear giving rise to lower mass fragments with less hyperons being attached to them. This establishes the occurrence of phase transition in hyper-fragments, a phenomenon which has already been observed in case of ordinary non strange fragments.\\

{\bf {\it Acknowledgement:-}}
The authors are gratefully acknowlege important discussions with Dr. D. N. Basu of VECC.

\end{document}